\documentclass[12pt]{article} 
\usepackage[utf8]{inputenc}
\usepackage{amsmath}
\usepackage{amsfonts}
\usepackage{amssymb}
\usepackage{amsmath}
\usepackage{amssymb}
\usepackage{amstext}
\usepackage{bm}
\usepackage{graphicx}
\usepackage{epstopdf}

\allowdisplaybreaks 

\begin{document}
\baselineskip= 0.225in
\parindent=0.35in

\begin{center}
\textbf{\large \\%
ASYMPTOTIC DERIVATION OF THE SIMPLIFIED P$_N$ EQUATIONS FOR NONCLASSICAL TRANSPORT WITH ANISOTROPIC SCATTERING
}

\vspace{0.2in}
\setlength{\baselineskip}{12pt}
\textbf{Robert Palmer and Richard Vasques}\\
\vspace{0.1in}
The Ohio State University, Department of Mechanical and Aerospace Engineering\\
201 W. 19th Avenue, Columbus, OH 43201\\
\vspace{0.1in}

palmer.462@osu.edu, vasques.4@osu.edu \\
\vspace{0.25in}
\end{center}

\vspace {0.25in} 

\section{Introduction}

\hspace{0.25in}

An accurate model of particle transport through scattering and absorbing media is necessary for the understanding of many phenomena in nuclear engineering and physics.  In classical transport theory, the particle flux is attenuated exponentially as particles move through a homogeneous medium.  In such a medium, the distribution of $s$, the free-path length of the particle since birth or creation, is exponential.  This exponential free-path length distribution arises because the scattering centers within the material are uncorrelated, or Poisson distributed.  Now, consider a system which consists of clumps of a homogeneous material within a matrix of another material with a significantly different macroscopic total cross section.  In this case, the locations of the scattering centers are spatially correlated (not Poisson distributed), and the distribution of particle free-path lengths is nonexponential, and therefore nonclassical.   In particular, if the ``clumps" of homogeneous material are suspended in a ``void", the random walk performed by each particle will be described by a L\`evy flight, and not by Brownian motion.  A L\`evy flight is characterized by a free-path length distribution with a heavy tail, which is nonclassical.  Such nonclassical transport occurs in neutron transport in pebble-bed reactors (in which the heterogeneous system is formed by fuel pebbles and coolant) \cite{vasques_1,vasques_2,vasques_3}, neutron transport in boiling water reators (in which the heterogeneous system is formed by water and steam bubbles) \cite{lahey_1}, in photon transport in clouds \cite{pfeilsticker_1,kostinski_1,buldyrev_1,kostinski_2,borovoi_1,kostinski_3,shaw_1,
davis_1,scholl_1,davis_2}, in Lorentz gases \cite{golse_1,marklof_1,marklof_2,marklof_3,marklof_4}, in light simulation in computer graphics \cite{deon_1,jarabo_1,bitterli_1}, and in a glass matrix embedded with high refractive index particles \cite{barthelemy_1}.  These are common, but not the only, ways in which systems can exhibit nonclassical transport.  

In diffusive regimes, defined as regimes which are optically thick and in which leakage out of the system is small and sources are weak, the simplified spherical harmonic equations ($SP_N$) are useful in solving transport problems.  However, the standard $SP_N$ equations may not accurately model nonclassical transport in diffusive systems.  To accurately model nonclassical transport in diffusive sytems, the theory of nonclassical transpor was created \cite{larsen_5}.  A nonclassical $SP_1$ equation with anisotropic scattering \cite{larsen_5} and the nonclassical $SP_N$ equations with isotropic scattering have been derived \cite{vasques_4}.  However, higher order nonclassical $SP_N$ equations which can model anisotropic scattering have not yet been determined.  This paper describes a method with which one can derive the nonclassical $SP_N$ equations with anisotropic scattering, and this paper uses this method to determine the first of these equations.  

The rest of this document is organized as follows: Section 2 will discuss the classical Boltzmann transport equation and introduce the nonclassical Boltzmann transport equation.  Section 3 will detail the scaling approach chosen to complete the asymptotic derivation.  Section 4 will present the asymptotic analysis employed to derive the nonclassical $SP_N$ equations with anisotropic scattering, and then the nonclassical $SP_1$ equation will be derived explicitly using this novel approach.  Finally, Section 5 will present a brief summary and discuss future work.

\pagebreak

\section{Particle Transport}

\hspace{0.25in}
In classical, steady-state, monoenergetic particle transport, the particle flux in a spatially homogeneous system can be modeled by the classical linear Boltzmann transport equation, which is given by
\begin{equation}
\bm \Omega \cdot \nabla \psi (\bm x, \bm \Omega) + \Sigma_t \psi (\bm x, \bm \Omega) =  \int_{4\pi} c \Sigma_t P(\bm \Omega \cdot \bm \Omega')  \psi (\bm x, \bm \Omega') d\Omega' + \frac{Q(\bm x)}{4\pi}.
\end{equation}
Here, $ \psi $ is the classical angular flux, $ \bm x = (x,y,z) $ is the particle's position, $ \bm \Omega = (\Omega_x,\Omega_y,\Omega_z) $ (with $\lvert \bm \Omega \rvert = 1$) is the particle's direction of flight, $ \Sigma_t $ is the total cross section, $ c $ is the scattering ratio, $ P(\bm \Omega \cdot \bm \Omega') $ is the distribution of particles with direction of flight $ \bm \Omega' $ which scatter into direction of flight $ \bm \Omega $, and $ Q $ is an interior isotropic particle source.

In spatially homogeneous media, the distribution of the particle free-path $ s $ is given by
\begin{equation}
p(s) = \Sigma_t e^{-\Sigma_t s}.
\end{equation}
This distribution is valid if the locations of the scattering centers are uncorrelated.  In nonclassical particle transport, the total cross section $ \Sigma_t $ is no longer treated as independent of the free-path.  Instead, it is modeled as a function of $ s $.  The ensemble-averaged total cross section $ \Sigma_t(s) $, defined by, $ \Sigma_t(s) ds = $ the probability (ensemble-averaged over all physical realizations) that a particle, scattered or born at any point $ \bm x $, and traveling in any direction $ \bm \Omega $ will experience a collision between $ \bm x + s \bm \Omega $ and $ \bm x + (s + ds) \bm \Omega $, is known.  (For problems in general random media,  $ \Sigma_t(s) $ depends also on $ \bm x $ and $ \bm \Omega $.  In this work, the statistics are assumed to be homogeneous and independent of the direction of flight, in which case $ \Sigma_t $ depends only on $ s $.)  The $ s $-dependent total cross section is related to the particle free-path distribution by \cite{larsen_5}
\begin{equation}
p(s) = \Sigma_t(s) e^{-\int_0^{s} \Sigma_t(s')ds'}.
\end{equation}
We observe that if the total cross section is constant with respect to $ s $, Equation (3) reduces to the exponential distribution given by Equation (2).  Solving Equation (3) for $\Sigma_t(s)$, \cite{larsen_5}
\begin{equation}
 \Sigma_t(s) = \frac{p(s)}{1 - \int_0^s p(s')ds'}.
\end{equation}
Raw moments of the distribution can be calculated using
\begin{equation}
\langle s^m \rangle = \int_0^\infty s^m p(s) ds .
\end{equation}
Similar to Equation (3), if the total cross section is independent of $s$, then these raw moments of the free-path length distribution are given by $\langle s^m \rangle = m! \Sigma_t^{-m}$.

Assuming steady-state, monoenergetic transport with an isotropic internal source, Equation (1) is modified to produce the nonclassical Boltzmann transport equation \cite{larsen_5}.  It differs from Equation (1) in its dependence upon the free-path length variable $ s $.
\begin{equation}
\frac{\partial}{\partial s} \hat{\psi}(\bm x, \bm \Omega, s) + \bm \Omega \cdot \nabla \hat{\psi} (\bm x, \bm \Omega, s) + \Sigma_t (s) \hat{\psi} (\bm x, \bm \Omega, s) \nonumber
\end{equation}
\begin{equation}
= \delta (s) \left[ \int_{4\pi} \int_0^{\infty} c \Sigma_t (s') P(\bm \Omega \cdot \bm \Omega')  \hat{\psi} (\bm x, \bm \Omega', s') ds'd\Omega' + \frac{Q(\bm x)}{4\pi} \right].
\end{equation}
Here, $ \hat{\psi} (\bm x, \bm \Omega, s) $ is the nonclassical angular flux.  The right side of Equation 6 contains the $ \delta (s) $ since a particle's free-path length equals zero after it scatters or is generated from the source $ Q $.  Also, the distribution of particles moving in direction $ \bm \Omega' $ that scatter into direction of flight $ \bm \Omega $ is given by the Legendre polynomial expansion
\begin{equation}
P(\bm \Omega \cdot \bm \Omega')  =  \sum_{m=0}^{\infty}\frac{2m+1}{4\pi}a_m P_m(\bm \Omega \cdot \bm \Omega'),
\end{equation}
where $P_m$ is the $m$th order Legendre polynomial and $a_m$ is the $m$th order Legendre polynomial expansion coefficient with $a_0 = 1$ and $a_1 = \bar{\mu}_0$, the mean scattering cosine.  Finally, the classical angular flux $ \psi (\bm x, \bm \Omega) $ can be recovered form the solution of Equation (6) by integrating over $ s $:
\begin{equation}
\psi (\bm x, \bm \Omega) = \int_0^\infty \hat{\psi} (\bm x, \bm \Omega, s) ds.
\end{equation}
Equation (6) is similar in form to the time-dependent monoenergetic, anisotropic Boltzmann transport equation.  We can therefore rewrite it in ``initial value" form as a system of two equations given by
\begin{equation}
\frac{\partial}{\partial s} \hat{\psi}(\bm x, \bm \Omega, s) + \bm \Omega \cdot \nabla \hat{\psi} (\bm x, \bm \Omega, s) + \Sigma_t (s) \hat{\psi} (\bm x, \bm \Omega, s) = 0, \hspace{0.1in} s > 0 \nonumber
\end{equation}
and
\begin{equation}
 \hat{\psi}(\bm x, \bm \Omega, 0) = \int_{4\pi} \int_0^{\infty} c \Sigma_t (s') P(\bm \Omega \cdot \bm \Omega')  \hat{\psi} (\bm x, \bm \Omega', s') ds'd\Omega' + \frac{Q(\bm x)}{4\pi}.
\end{equation}

The nonclassical simplified $P_N$ equations are $SP_N$ equations used in nonclassical transport regimes to solve Equation (6).  Larsen and Vasques \cite{larsen_5} begin with Equation (6), and assuming a globally diffusive regime, use an asymptotic analysis to produce the nonclassical diffusion ($SP_1$) equation with anisotropic scattering, given by
\begin{equation}
-\frac{1}{3} \left[ \frac{\langle s^2 \rangle}{2 \langle s \rangle} + \frac{c \bar{\mu}_0}{1-c \bar{\mu}_0} \langle s \rangle \right] \nabla^2 \Phi(\bm x) + \frac{1-c}{\langle s \rangle} \Phi(\bm x) = Q(\bm x),
\end{equation}
where $\Phi(\bm x) = \int_{4\pi} \psi(\bm x, \bm \Omega) d\Omega$ is the classical scalar flux.  In this equation, information about the nonclassical nature of the medium is contained in the moments of $s$, and information about the anisotropic scattering produced by the medium is contained in the mean scattering cosine.  If the free-path length distribution is exponential and therefore obeys Equation (2) ($s$-independent), then Equation (10) reduces to the classical diffusion equation with anisotropic scattering
\begin{equation}
-\frac{1}{3 \Sigma_t (1 - c \bar{\mu}_0)} \nabla^2 \Phi(\bm x) + \Sigma_a \Phi(\bm x) = Q(\bm x),
\end{equation}
where $\Sigma_a =  \Sigma_t (1 - c)$.  However, the asymptotic method employed in \cite{larsen_5} is unable to produce higher-order nonclassical diffusion equations.

Vasques and Slaybaugh \cite{vasques_4} used a different asymptotic approach to derive a method to generate the nonclassical $SP_N$ equations with isotropic scattering.  In their work, they explicitly derived the nonclassical $SP_1$ equation,
\begin{equation}
-\frac{1}{6} \frac{\langle s^2 \rangle}{\langle s \rangle} \nabla^2 \Phi(\bm x) + \frac{1-c}{\langle s \rangle} \Phi(\bm x) = Q(\bm x).
\end{equation}

A goal of this work is to generalize the nonclassical $SP_N$ equations with isotropic scattering so that they can accurately simulate anisotropic diffusion.  First, we will describe the scaling used in the asymptotic analysis to derive the nonclassical $SP_1$ equation with anisotropic scattering.

\pagebreak

\section{Scaling Approach}

\hspace{0.25in}
The following scaling approach used is identical to that used by Vasques and Slaybaugh \cite{vasques_4}.  Defining $ 0 < \varepsilon \ll 1 $, the following scaling relationships are applied.
\begin{equation}
\Sigma_t(s) = \frac{\sigma(s/\varepsilon)}{\varepsilon}.
\end{equation}
\begin{equation}
Q(\bm x) = \varepsilon q(\bm x).
\end{equation}
\begin{equation}
1-c = \varepsilon^2 \kappa.
\end{equation}
Here, $ \kappa $ and $ q $ are $ O(1) $.  These choices result in the scaled moment
\begin{equation}
\langle s^m \rangle = \varepsilon^m \int_0^{\infty} \left( \frac{s}{\varepsilon} \right)^m \frac{\sigma(s/\varepsilon)}{\varepsilon} e^{-\int_0^s \frac{\sigma(s'/\varepsilon)}{\varepsilon} ds'} ds = \varepsilon^m \langle s^m \rangle_\varepsilon,
\end{equation}
which is $ O(1) $.  These scaling choices imply that the system is optically thick and that the influence of absorption and sources are small comparable to that of scattering.  We now define
\begin{equation}
 \hat{\psi}(\bm x, \bm \Omega, \varepsilon s) \equiv  \psi_\varepsilon (\bm x, \bm \Omega, s).
\end{equation}
We combine the scaling relationships given by Equation (13) through (15), and we can now write Equation (9) in a manner which is satisfied by Equation (17) as
\begin{equation}
\frac{\partial}{\partial s} \psi_\varepsilon (\bm x, \bm \Omega, s) + \varepsilon \bm \Omega \cdot \nabla \psi_\varepsilon (\bm x, \bm \Omega, s) + \sigma (s) \psi_\varepsilon (\bm x, \bm \Omega, s) = 0,  s > 0.
\end{equation}
\begin{equation}
 \psi_\varepsilon (\bm x, \bm \Omega, 0) = \int_{4\pi} \int_0^{\infty} (1 - \varepsilon^2 \kappa) \sigma (s') P(\bm \Omega \cdot \bm \Omega') \psi_\varepsilon (\bm x, \bm \Omega', s') ds'd\Omega' + \varepsilon \frac{q(\bm x)}{4\pi}.
\end{equation}
Then we define $\Psi$, which satisfies
\begin{equation}
\psi_\varepsilon (\bm x, \bm \Omega, s) \equiv \Psi (\bm x, \bm \Omega, s) \frac{e^{-\int_0^\infty \frac{\sigma (s'/\varepsilon)}{\varepsilon} ds' }}{\varepsilon \langle s \rangle_\varepsilon}.
\end{equation}
where $ \langle s \rangle_\varepsilon $ is the first moment of $ s $ given by Equation (16).  Then, Equations (18) and (19) become
\begin{equation}
\frac{\partial}{\partial s} \Psi (\bm x, \bm \Omega, s) + \varepsilon \bm \Omega \cdot \nabla \Psi (\bm x, \bm \Omega, s) = 0,  s > 0.
\end{equation}
\begin{equation}
 \Psi (\bm x, \bm \Omega, 0) = \int_{4\pi} \int_0^{\infty} (1 - \varepsilon^2 \kappa) p(s') P(\bm \Omega \cdot \bm \Omega') \Psi (\bm x, \bm \Omega', s') ds'd\Omega' + \varepsilon^2 \langle s \rangle_\varepsilon \frac{q(\bm x)}{4\pi},
\end{equation}
where $p(s)$ is given by Equation (3).  Integrating Equation (21) from 0 to $s$ and then combining this result with Equation (22) yields
\begin{equation}
\Psi (\bm x, \bm \Omega, s) + \varepsilon \bm \Omega \cdot \nabla \int_0^s \Psi (\bm x, \bm \Omega, s') ds' \nonumber
\end{equation}
\begin{equation}
= \int_{4\pi} \int_0^{\infty} (1 - \varepsilon^2 \kappa) p(s') P(\bm \Omega \cdot \bm \Omega') \Psi (\bm x, \bm \Omega', s') ds'd\Omega' + \varepsilon^2 \langle s \rangle_\varepsilon \frac{q(\bm x)}{4\pi}.
\end{equation}

\pagebreak

\section{Procedure to Derive the Nonclassical $SP_N$ Equations}

\hspace{0.25in}
This section describes the procedure developed to derive the nonclassical $SP_N$ equations.  The asymptotic analysis used is a generalization of the asymptotic analyses employed by \cite{vasques_4} and \cite{larsen_6}.  Subsection 4.1 presents an asymptotic analysis of Equation (23).  In Subsection 4.2, we apply this asymptotic analysis to explicitly derive the nonclassical $SP_1$ equation with anisotropic scattering. 

\subsection{Asymptotic Analysis}

\hspace{0.25in}
Now we define the following:
\begin{equation}
\varphi (\bm x,s) \equiv \int_{4 \pi} \Psi (\bm x, \bm \Omega, s) d \Omega.
\end{equation}
\begin{equation}
\Im \Psi (\bm x, \bm \Omega, s) \equiv \frac{1}{4 \pi} \int_{4 \pi} \Psi (\bm x, \bm \Omega, s) d \Omega.
\end{equation}
We operate on Equation (48) using $ \Im $. Since $ P(\bm \Omega \cdot \bm \Omega') $ is defined as a Legendre polynomial expansion given by Equation (7), all terms of $ P(\bm \Omega \cdot \bm \Omega') $ integrate to zero (due to orthogonality) except for the zeroth term, $a_0/(4\pi)$ with $a_0 = 1$.  This results in
\begin{equation}
\frac{\varphi (\bm x,s)}{4 \pi} + \varepsilon \Im \bm \Omega \cdot \nabla \int_0^s \Psi (\bm x, \bm \Omega, s') ds' = \frac{1}{4 \pi} \int_0^\infty (1 - \varepsilon^2 \kappa) p(s') \varphi (\bm x,s')ds' + \varepsilon^2 \langle s \rangle_\varepsilon \frac{q(\bm x)}{4\pi}.
\end{equation}
Next, we subtract Equation (26) from Equation (23) to get 
\begin{equation}
\varepsilon (I - \Im) \bm \Omega \cdot \nabla \int_0^s \Psi (\bm x, \bm \Omega, s') ds' + \Psi (\bm x, \bm \Omega, s) - \frac{\varphi (\bm x,s)}{4 \pi} \nonumber
\end{equation}
\begin{equation}
 = \int_{4\pi} \int_0^{\infty} (1 - \varepsilon^2 \kappa) p(s') \sum_{m=1}^{\infty}\frac{2m+1}{4\pi}a_m P_m(\bm \Omega \cdot \bm \Omega') \Psi (\bm x, \bm \Omega', s') ds'd\Omega'.
\end{equation}
Equations (26) and (27) provide two equations for the two unknowns, $ \Psi (\bm x, \bm \Omega, s) $ and $ \varphi (\bm x,s) $.  We note that the sum in Equation (27) now begins at 1 since the zeroth term, which is the isotropic term, cancels in subtraction.  Now, we define the operator $ \jmath $,
\begin{equation}
\jmath \Psi (\bm x, \bm \Omega, s) \equiv \Psi (\bm x, \bm \Omega, s) \nonumber
\end{equation}
\begin{equation}
 - \int_{4\pi} \int_0^{\infty} (1 - \varepsilon^2 \kappa) p(s') \sum_{m=1}^{\infty}\frac{2m+1}{4\pi}a_mP_m(\bm \Omega \cdot \bm \Omega') \Psi (\bm x, \bm \Omega', s') ds'd\Omega',
\end{equation}
which allows us to rewrite Equation (27) in the more compact form
\begin{equation}
\jmath \Psi (\bm x, \bm \Omega, s) + \varepsilon (I - \Im) \bm \Omega \cdot \nabla \int_0^s \Psi (\bm x, \bm \Omega, s') ds' = \frac{\varphi (\bm x,s)}{4 \pi}.
\end{equation}
Operating on Equation (29) by $ \jmath^{-1} $ yields
\begin{equation}
\Psi (\bm x, \bm \Omega, s) + \varepsilon \jmath^{-1} (I - \Im) \bm \Omega \cdot \nabla \int_0^s \Psi (\bm x, \bm \Omega, s') ds' = \jmath^{-1} \left[ \frac{\varphi (\bm x,s)}{4 \pi} \right],
\end{equation}
and rewriting the left side in operator form results in
\begin{equation}
 \left[ I + \varepsilon \jmath^{-1} (I - \Im) \bm \Omega \cdot \nabla \int_0^s (\cdot) ds'   \right] \Psi (\bm x, \bm \Omega, s) = \frac{1}{4 \pi} \jmath^{-1} \varphi (\bm x,s).
\end{equation}
Inverting the operator on the left side of Equation (31) yields
\begin{equation}
\Psi (\bm x, \bm \Omega, s) = \frac{1}{4 \pi} \left[ I + \varepsilon \jmath^{-1} (I - \Im) \bm \Omega \cdot \nabla \int_0^s (\cdot) ds'   \right]^{-1} \jmath^{-1} \varphi (\bm x,s).
\end{equation}
Expanding the inverse operator in Equation (32) in a Taylor series about $ \varepsilon = 0 $ produces the following result for $ \Psi (\bm x, \bm \Omega, s) $:
\begin{equation}
\Psi (\bm x, \bm \Omega, s) = \frac{1}{4 \pi} \sum_{i=0}^\infty (-1)^i \varepsilon^i \left[ \jmath^{-1} (I - \Im) \bm \Omega \cdot \nabla \int_0^s (\cdot) ds' \right]^i \jmath^{-1} \varphi (\bm x,s).
\end{equation}
To proceed, either an exact expression or a Taylor expansion for $ \jmath^{-1} $ must be determined.  The operator $ \jmath $ is modified from Equation (28), 
\begin{equation}
\jmath \Psi (\bm x, \bm \Omega, s) \equiv \Psi (\bm x, \bm \Omega, s) - \int_{4\pi} \int_0^{\infty} (1 - \varepsilon^2 \kappa) p(s') \sum_{n=0}^{\infty}\frac{2n+1}{4\pi} b_n P_n(\bm \Omega \cdot \bm \Omega') \Psi (\bm x, \bm \Omega', s') ds'd\Omega',
\end{equation}
where $P_n$ is the $n$th order Legendre polynomial and $b_n$ is a modified $n$th order Legendre polynomial expansion coefficient where $b_0 = 0$ and $b_n \geq 1$ is a regular $n$th order Legendre polynomial expansion coefficient.  To generate the inverse function, we first use the Addition Theorem of spherical harmonic functions to rewrite $ P_n(\bm \Omega \cdot \bm \Omega') $ as
\begin{equation}
P_n(\bm \Omega \cdot \bm \Omega') = \frac{4 \pi}{2n+1}  \sum_{m=-n}^{n} Y_n^m (\bm \Omega) \bar{Y}_n^m (\bm \Omega') 
\end{equation}
where $  Y_n^m (\bm \Omega) $ is a spherical harmonic function and $ \bar{Y}_n^m (\bm \Omega')  $ is its complex conjugate.  
We now rewrite $ \jmath \Psi (\bm x, \bm \Omega, s) $ as
\begin{equation}
\jmath \Psi (\bm x, \bm \Omega, s) \nonumber
\end{equation}
\begin{equation}
 = \Psi (\bm x, \bm \Omega, s) - \int_{4\pi} \int_0^{\infty} (1 - \varepsilon^2 \kappa) p(s') \sum_{n=0}^{\infty}\frac{2n+1}{4\pi} b_n \left(  \frac{4 \pi}{2n+1}  \sum_{m=-n}^{n} Y_n^m (\bm \Omega) \bar{Y}_n^m (\bm \Omega')  \right) \Psi (\bm x, \bm \Omega', s') ds'd\Omega'.
\end{equation}
We can also write $  \Psi (\bm x, \bm \Omega, s) $ as an expansion in spherical harmonics as
\begin{equation}
\Psi (\bm x, \bm \Omega, s) = \sum_{n=0}^{\infty} \sum_{m=-n}^{n} \chi_n^m (\bm x, s) Y_n^m (\bm \Omega) 
\end{equation}
where
\begin{equation}
\chi_n^m (\bm x, s) = \int_{4\pi} \Psi (\bm x, \bm \Omega', s) \bar{Y}_n^m (\bm \Omega') d\Omega'.
\end{equation}
Then Equation (36) becomes
\begin{equation}
\jmath \Psi (\bm x, \bm \Omega, s) \nonumber
\end{equation}
\begin{equation}
= \sum_{n=0}^{\infty} \sum_{m=-n}^{n} \left[ \chi_n^m (\bm x, s) Y^m (\bm \Omega')  - \int_0^{\infty} (1 - \varepsilon^2 \kappa) p(s') b_n Y_n^m (\bm \Omega) \left( \int_{4 \pi} \bar{Y}_n^m (\bm \Omega') \Psi (\bm x, \bm \Omega', s') d \Omega' \right) ds'\right].
\end{equation}
The angular integral within parentheses on the right side of Equation (39) equals $ \chi_n^m (\bm x, s) $, so we can rewrite Equation (39) as
\begin{equation}
\jmath \Psi (\bm x, \bm \Omega, s) \nonumber
\end{equation}
\begin{equation}
= \sum_{n=0}^{\infty} \sum_{m=-n}^{n} Y_n^m (\bm \Omega) \left[ \chi_n^m (\bm x, s) - (1 - \varepsilon^2 \kappa) b_n \int_0^{\infty} p(s') \chi_n^m (\bm x, s') ds'\right].
\end{equation}
Next, we define for any function $ f(\bm x, s) $,
\begin{equation}
L_n f(\bm x, s) \equiv f(\bm x, s) - (1 - \varepsilon^2 \kappa) b_n \int_0^{\infty} p(s') f(\bm x, s') ds',
\end{equation}
so that we can write $ \jmath \Psi (\bm x, \bm \Omega, s) $ more compactly as
\begin{equation}
\jmath \Psi (\bm x, \bm \Omega, s) =  \sum_{n=0}^{\infty} \sum_{m=-n}^{n} Y_n^m (\bm \Omega) L_n \chi_n^m (\bm x, s).
\end{equation}
Now, we claim that 
\begin{equation}
\jmath^{-1} \Psi (\bm x, \bm \Omega, s) =  \sum_{n=0}^{\infty} \sum_{m=-n}^{n} Y_n^m (\bm \Omega) L_n^{-1} \chi_n^m (\bm x, s).
\end{equation}
So, we must determine $ L_n^{-1} f(\bm x, s) $.  
First, we call
\begin{equation}
g(\bm x, s) = L_n f(\bm x, s),
\end{equation}
so that we want to solve $ f(\bm x, s) = L_n^{-1} g(\bm x, s) $.
First, multiply Equation (44) by $ p(s) $ (and using Equation (41)),
\begin{equation}
p(s) f(\bm x, s) - (1 - \varepsilon^2 \kappa) b_n p(s) \int_0^{\infty} p(s') f(\bm x, s') ds' = p(s) g(\bm x, s),
\end{equation}
and then we operate on this by $ \int_0^{\infty} (\cdot) ds $ to get
\begin{equation}
\int_0^{\infty} p(s) f(\bm x, s) ds - (1 - \varepsilon^2 \kappa) b_n \int_0^{\infty} p(s') f(\bm x, s') ds' = \int_0^{\infty} p(s) g(\bm x, s) ds.
\end{equation}
We solve this equation for $ \int_0^{\infty} p(s) f(\bm x, s) ds $ to arrive at
\begin{equation}
\int_0^{\infty} p(s) f(\bm x, s) ds = \frac{1}{1 - (1 - \varepsilon^2 \kappa) b_n} \int_0^{\infty} p(s) g(\bm x, s) ds.
\end{equation}
Now, $ L_n^{-1} g(\bm x, s) $ becomes
\begin{equation}
 f(\bm x, s) = L_n^{-1} g(\bm x, s) = g(\bm x, s) + \frac{(1 - \varepsilon^2 \kappa) b_n}{1-(1 - \varepsilon^2 \kappa) b_n} \int_0^{\infty} p(s) g(\bm x, s) ds.
\end{equation}
Finally, to validate Equation (43), we will show that if $ L_n^{-1} \left[ L_n g(\bm x, s) \right] = 1 $, then $ \jmath^{-1} \left[ \jmath  \Psi (\bm x, \bm \Omega, s) \right] = 1 $.  We proceed as follows:
\begin{equation}
\jmath^{-1} \left[ \jmath  \Psi (\bm x, \bm \Omega, s) \right] =  \sum_{n=0}^{\infty} \sum_{m=-n}^{n} Y_n^m (\bm \Omega) L_n^{-1} \left[ \int_{4\pi} \left( \jmath \Psi (\bm x, \bm \Omega', s) \right) \bar{Y}_n^m (\bm \Omega') d\Omega' \right] \nonumber
\end{equation}
\begin{equation}
= \sum_{n=0}^{\infty} \sum_{m=-n}^{n} Y_n^m (\bm \Omega) L_n^{-1} \left[ \int_{4\pi} \bar{Y}_n^m (\bm \Omega') \left(  \sum_{n'=0}^{\infty} \sum_{m'=-n}^{n'} Y_{n'}^{m'} (\bm \Omega') L_n \left[ \chi_{n'}^{m'} (\bm x, s) \right] \right) \right] d\Omega' \nonumber
\end{equation}
\begin{equation}
= \sum_{n=0}^{\infty} \sum_{m=-n}^{n} \sum_{n'=0}^{\infty} \sum_{m'=-n}^{n'} Y_n^m (\bm \Omega) L_n^{-1} \left[ \int_{4\pi} \bar{Y}_n^m (\bm \Omega')  Y_{n'}^{m'} (\bm \Omega')  d\Omega' \right] L_n \left[ \chi_{n'}^{m'} (\bm x, s) \right].
\end{equation}
From the orthogonality of the spherical harmonic functions, 
\begin{equation}
 \int_{4\pi} \bar{Y}_n^m (\bm \Omega')  Y_{n'}^{m'} (\bm \Omega')  d\Omega' = \delta_{n,n'} \delta_{m,m'} = 1
\end{equation}
when $ n = n' $ and $ m = m' $ and equals zero otherwise.  Therefore, the primed sums in Equation (49) disappear, and 
\begin{equation}
\jmath^{-1} \left[ \jmath  \Psi (\bm x, \bm \Omega, s) \right] = \sum_{n=0}^{\infty} \sum_{m=-n}^{n} Y_n^m (\bm \Omega) L_n^{-1} \left( L_n \left[ \chi_{n'}^{m'} (\bm x, s) \right] \right).
\end{equation}
Since 
\begin{equation}
L_n^{-1} \left( L_n \left[ \chi_{n'}^{m'} (\bm x, s) \right] \right) = 1,
\end{equation}
then 
\begin{equation}
\jmath^{-1} \left[ \jmath  \Psi (\bm x, \bm \Omega, s) \right] = \sum_{n=0}^{\infty} \sum_{m=-n}^{n} Y_n^m (\bm \Omega) \chi_{n'}^{m'} (\bm x, s) = \Psi (\bm x, \bm \Omega, s).
\end{equation}
Therefore, Equation (43) is the inverse function for Equation (34).  So, the inverse function of Equation (28) is

\begin{equation}
\jmath^{-1} \Psi (\bm x, \bm \Omega, s) =  \sum_{n=0}^{\infty} \sum_{m=-n}^{n} Y_n^m (\bm \Omega) L_n^{-1} \chi_n^m (\bm x, s), 
\end{equation}
where 
\begin{equation}
L_n^{-1} g(\bm x, s) = g(\bm x, s) + \frac{(1 - \varepsilon^2 \kappa) b_n}{1-(1 - \varepsilon^2 \kappa) b_n} \int_0^{\infty} p(s') g(\bm x, s') ds',
\end{equation}
and
\begin{equation}
\chi_n^m (\bm x, s) = \int_{4\pi} \Psi (\bm x, \bm \Omega', s) \bar{Y}_n^m (\bm \Omega') d\Omega'.
\end{equation}
Also, as previously stated, the $ b_n $ coefficients are modified Legendre polynomial expansion coefficients with
\begin{equation}
b_0 = 0, \hspace{0.1in} b_n = a_n, \hspace{0.1in} n \geq 1
\end{equation}
The nonclassical flux $ \varphi (\bm x,s) $ in Equation (33) is not a function of $ \bm \Omega $, so it can be shown that
\begin{equation}
\jmath \varphi (\bm x,s) = \jmath^{-1} \varphi (\bm x,s) = \varphi (\bm x,s).
\end{equation}
So, Equation (58) can be simplified to the following equation:
\begin{equation}
\Psi (\bm x, \bm \Omega, s) = \frac{1}{4 \pi} \sum_{i=0}^\infty (-1)^i \varepsilon^i \left[ \jmath^{-1} (I - \Im) \bm \Omega \cdot \nabla \int_0^s (\cdot) ds' \right]^i \varphi (\bm x,s).
\end{equation}
From this equation, we begin the process of evaluating $ \Psi (\bm x, \bm \Omega, s) $.  For $ i = 0 $,
\begin{equation}
\Psi (\bm x, \bm \Omega, s) = \frac{1}{4 \pi} \varphi (\bm x,s).
\end{equation}
For $ i = 1 $, consider the term within the brackets in Equation (59).  It can be shown that 
\begin{equation}
\Im \bm \Omega \cdot \nabla \int_0^s \varphi (\bm x,s') ds' = 0,
\end{equation}
since this is an odd function of $ \bm \Omega $ under the operator $ \Im $.  Continuing with the term in brackets in Equation (59), the operator $\jmath^{-1} $ and its argument must be evaluated.  Since our current goal is to produce $ SP_1 $,  we choose to truncate the sum in Equation (59) at $ i = 1 $.  Applying Equation (54) results in 
\begin{equation}
\jmath^{-1} \bm \Omega \cdot \nabla \int_0^s \varphi (\bm x,s') ds' =  \sum_{n=0}^{1} \sum_{m=-1}^{1} Y_n^m (\bm \Omega) L_n^{-1} \int_{4\pi} \bm \Omega' \cdot \nabla \int_0^s \varphi (\bm x,s') ds' \bar{Y}_n^m (\bm \Omega') d\Omega', 
\end{equation}
For $ i = 1 $ we expand to get 
\begin{equation}
\jmath^{-1} \bm \Omega \cdot \nabla \int_0^s \varphi (\bm x,s') ds' \nonumber
\end{equation}
\begin{equation}
= Y_0^0 (\bm \Omega) L_0^{-1} \left[ \int_{4\pi} \bar{Y}_0^0 (\bm \Omega') \left(\bm \Omega' \cdot \nabla \int_0^s \varphi (\bm x,s') ds' \right) d\Omega' \right] \nonumber
\end{equation}
\begin{equation}
+ Y_1^{-1} (\bm \Omega) L_1^{-1} \left[ \int_{4\pi} \bar{Y}_1^{-1} (\bm \Omega') \left(\bm \Omega' \cdot \nabla \int_0^s \varphi (\bm x,s') ds' \right) d\Omega' \right] \nonumber
\end{equation}
\begin{equation}
+ Y_1^0 (\bm \Omega) L_1^{-1} \left[ \int_{4\pi} \bar{Y}_1^0 (\bm \Omega') \left(\bm \Omega' \cdot \nabla \int_0^s \varphi (\bm x,s') ds' \right) d\Omega' \right] \nonumber
\end{equation}
\begin{equation}
+ Y_1^1 (\bm \Omega) L_1^{-1} \left[ \int_{4\pi} \bar{Y}_1^1 (\bm \Omega') \left(\bm \Omega' \cdot \nabla \int_0^s \varphi (\bm x,s') ds' \right) d\Omega' \right].
\end{equation}
We will use the following relationship for spherical harmonic functions.
\begin{equation}
\bar{Y}_n^m (\bm \Omega) = (-1)^m Y_n^{-m} (\bm \Omega),
\end{equation}
and the necessary spherical harmonic functions
\begin{equation}
Y_0^0 (\bm \Omega) = \frac{1}{\sqrt{4 \pi}},
\end{equation}
\begin{equation}
Y_1^{-1} (\bm \Omega) = \sqrt{\frac{3}{8 \pi}} (\Omega_x - i \Omega_y),
\end{equation}
\begin{equation}
Y_1^0 (\bm \Omega) = \sqrt{\frac{3}{4 \pi}} \Omega_z,
\end{equation}
and
\begin{equation}
Y_1^1 (\bm \Omega) = - \sqrt{\frac{3}{8 \pi}} (\Omega_x + i \Omega_y).
\end{equation}
Evaluating the first term on the right side of Equation (63) results in
\begin{equation}
Y_0^0 (\bm \Omega) L_0^{-1} \left[ \int_{4\pi} (-1)^0 Y_0^0 (\bm \Omega') \left(\bm \Omega' \cdot \nabla \int_0^s \varphi (\bm x,s') ds' \right) d\Omega' \right] \nonumber
\end{equation}
\begin{equation}
= \frac{1}{\sqrt{4 \pi}} L_0^{-1} \left[ \int_{4\pi} \frac{1}{\sqrt{4 \pi}} \left(\bm \Omega' \cdot \nabla \int_0^s \varphi (\bm x,s') ds' \right) d\Omega' \right] = 0.
\end{equation}
This is true since we are integrating an odd function over the entire unit sphere.
Evaluating the second term on the right side of Equation (63) results in
\begin{equation}
Y_1^{-1} (\bm \Omega) L_1^{-1} \left[ \int_{4\pi} \bar{Y}_1^{-1} (\bm \Omega') \left(\bm \Omega' \cdot \nabla \int_0^s \varphi (\bm x,s') ds' \right) d\Omega' \right] \nonumber
\end{equation}
\begin{equation}
= \sqrt{\frac{3}{8 \pi}} (\Omega_x - i \Omega_y) L_n^{-1} \left[ \int_{4\pi} \sqrt{\frac{3}{8 \pi}} (\Omega'_x + i \Omega'_y) \left(\bm \Omega' \cdot \nabla \int_0^s \varphi (\bm x,s') ds' \right) d\Omega' \right].
\end{equation}
We combine and integrate the terms within brackets to yield
\begin{equation}
= \frac{3}{8 \pi} (\Omega_x - i \Omega_y) L_n^{-1} \left[ \left( \frac{4 \pi}{3} \frac{\partial}{\partial x}\int_0^s \varphi (\bm x,s') ds' + i \frac{4 \pi}{3} \frac{\partial}{\partial y}\int_0^s \varphi (\bm x,s') ds' \right) \right],
\end{equation}
and
\begin{equation}
= \frac{1}{2} (\Omega_x - i \Omega_y) L_n^{-1} \left[ \left( \frac{\partial}{\partial x} + i \frac{\partial}{\partial y} \right) \int_0^s \varphi (\bm x,s') ds' \right].
\end{equation}
Now employing the definition of the operator $ L_1^{-1} $ yields
\begin{equation}
= \frac{1}{2} (\Omega_x - i \Omega_y) \left[ \left( \frac{\partial}{\partial x} + i \frac{\partial}{\partial y} \right) \int_0^s \varphi (\bm x,s') ds'  \right] \nonumber
\end{equation}
\begin{equation}
+ \frac{1}{2} (\Omega_x - i \Omega_y) \left[ \frac{(1 - \varepsilon^2 \kappa) b_1}{1-(1 - \varepsilon^2 \kappa) b_1} \int_0^{\infty} p(s'') \left( \frac{\partial}{\partial x} + i \frac{\partial}{\partial y} \right) \int_0^{s''} \varphi (\bm x,s') ds' ds'' \right]
\end{equation}
\begin{equation}
= \frac{1}{2} (\Omega_x - i \Omega_y) \left( \frac{\partial}{\partial x} + i \frac{\partial}{\partial y} \right)  \left[ \int_0^s \varphi (\bm x,s') ds' + \frac{(1 - \varepsilon^2 \kappa) b_1}{1-(1 - \varepsilon^2 \kappa) b_1} \int_0^{\infty} p(s'')  \int_0^{s''} \varphi (\bm x,s') ds' ds'' \right].
\end{equation}
Evaluating the third term on the right side of Equation (63) yields
\begin{equation}
Y_1^0 (\bm \Omega) L_1^{-1} \left[ \int_{4\pi} \bar{Y}_1^0 (\bm \Omega') \left(\bm \Omega' \cdot \nabla \int_0^s \varphi (\bm x,s') ds' \right) d\Omega' \right] \nonumber
\end{equation}
\begin{equation}
= \sqrt{\frac{3}{4 \pi}} \Omega_z L_n^{-1} \left[ \int_{4\pi} \sqrt{\frac{3}{4 \pi}} \Omega'_z \left(\bm \Omega' \cdot \nabla \int_0^s \varphi (\bm x,s') ds' \right) d\Omega' \right].
\end{equation}
Combining and integrating the terms within brackets results in
\begin{equation}
= \sqrt{\frac{3}{4 \pi}} \Omega_z L_n^{-1} \left[ \sqrt{\frac{3}{4 \pi}} \frac{4 \pi}{3} \frac{\partial}{\partial z}\int_0^s \varphi (\bm x,s') ds' \right]
\end{equation}
and
\begin{equation}
= \Omega_z L_n^{-1} \left[ \frac{\partial}{\partial z}\int_0^s \varphi (\bm x,s') ds' \right].
\end{equation}
Using the definition of $ L_1^{-1} $ results in
\begin{equation}
=\Omega_z \left[ \frac{\partial}{\partial z} \int_0^s \varphi (\bm x,s') ds' + \frac{(1 - \varepsilon^2 \kappa) b_1}{1-(1 - \varepsilon^2 \kappa) b_1} \frac{\partial}{\partial z} \int_0^{\infty} p(s'')  \int_0^{s''} \varphi (\bm x,s') ds' ds'' \right]
\end{equation}
\begin{equation}
=\Omega_z \frac{\partial}{\partial z} \left[ \int_0^s \varphi (\bm x,s') ds' + \frac{(1 - \varepsilon^2 \kappa) b_1}{1-(1 - \varepsilon^2 \kappa) b_1} \int_0^{\infty} p(s'')  \int_0^{s''} \varphi (\bm x,s') ds' ds'' \right].
\end{equation}
Evaluating the last term on the right side of Equation (63) gives us
\begin{equation}
Y_1^1(\bm \Omega) L_1^{-1} \left[ \int_{4\pi} \bar{Y}_1^1 (\bm \Omega') \left(\bm \Omega' \cdot \nabla \int_0^s \varphi (\bm x,s') ds' \right) d\Omega' \right] \nonumber
\end{equation}
\begin{equation}
= -\sqrt{\frac{3}{8 \pi}} (\Omega_x + i \Omega_y) L_n^{-1} \left[ \int_{4\pi} -\sqrt{\frac{3}{8 \pi}} (\Omega'_x - i \Omega'_y) \left(\bm \Omega' \cdot \nabla \int_0^s \varphi (\bm x,s') ds' \right) d\Omega' \right].
\end{equation}
Combining and integrating the terms within brackets results in
\begin{equation}
= \frac{3}{8 \pi} (\Omega_x + i \Omega_y) L_n^{-1} \left[ \left( \frac{4 \pi}{3} \frac{\partial}{\partial x}\int_0^s \varphi (\bm x,s') ds' - i \frac{4 \pi}{3} \frac{\partial}{\partial y}\int_0^s \varphi (\bm x,s') ds' \right) \right]
\end{equation}
and
\begin{equation}
= \frac{1}{2} (\Omega_x + i \Omega_y) L_n^{-1} \left[ \left( \frac{\partial}{\partial x} - i \frac{\partial}{\partial y} \right) \int_0^s \varphi (\bm x,s') ds' \right].
\end{equation}
Now employing the definition of the operator $ L_n^{-1} $ yields
\begin{equation}
= \frac{1}{2} (\Omega_x + i \Omega_y) \left[ \left( \frac{\partial}{\partial x} - i \frac{\partial}{\partial y} \right) \int_0^s \varphi (\bm x,s') ds'  \right] \nonumber
\end{equation}
\begin{equation}
+ \frac{1}{2} (\Omega_x + i \Omega_y) \left[ \frac{(1 - \varepsilon^2 \kappa) b_1}{1-(1 - \varepsilon^2 \kappa) b_1} \int_0^{\infty} p(s'') \left( \frac{\partial}{\partial x} - i \frac{\partial}{\partial y} \right) \int_0^{s''} \varphi (\bm x,s') ds' ds'' \right]
\end{equation}
\begin{equation}
= \frac{1}{2} (\Omega_x + i \Omega_y) \left( \frac{\partial}{\partial x} - i \frac{\partial}{\partial y} \right)  \left[ \int_0^s \varphi (\bm x,s') ds' + \frac{(1 - \varepsilon^2 \kappa) b_1}{1-(1 - \varepsilon^2 \kappa) b_1} \int_0^{\infty} p(s'')  \int_0^{s''} \varphi (\bm x,s') ds' ds'' \right].
\end{equation}
Combining all these terms, Equation (62) becomes
\begin{equation}
\jmath^{-1} \bm \Omega \cdot \nabla \int_0^s \varphi (\bm x,s') ds' \nonumber
\end{equation}
\begin{equation}
= \frac{1}{2} (\Omega_x - i \Omega_y) \left( \frac{\partial}{\partial x} + i \frac{\partial}{\partial y} \right)  \left[ \int_0^s \varphi (\bm x,s') ds' + \frac{(1 - \varepsilon^2 \kappa) b_1}{1-(1 - \varepsilon^2 \kappa) b_1} \int_0^{\infty} p(s'')  \int_0^{s''} \varphi (\bm x,s') ds' ds'' \right] \nonumber
\end{equation}
\begin{equation}
+ \Omega_z \frac{\partial}{\partial z} \left[ \int_0^s \varphi (\bm x,s') ds' + \frac{(1 - \varepsilon^2 \kappa) b_1}{1-(1 - \varepsilon^2 \kappa) b_1} \int_0^{\infty} p(s'')  \int_0^{s''} \varphi (\bm x,s') ds' ds'' \right] \nonumber
\end{equation}
\begin{equation}
 \frac{1}{2} (\Omega_x + i \Omega_y) \left( \frac{\partial}{\partial x} - i \frac{\partial}{\partial y} \right)  \left[ \int_0^s \varphi (\bm x,s') ds' + \frac{(1 - \varepsilon^2 \kappa) b_1}{1-(1 - \varepsilon^2 \kappa) b_1} \int_0^{\infty} p(s'')  \int_0^{s''} \varphi (\bm x,s') ds' ds'' \right].
\end{equation}
Simplifying gives us a more compact version,
\begin{equation}
\jmath^{-1} \bm \Omega \cdot \nabla \int_0^s \varphi (\bm x,s') ds' \nonumber
\end{equation}
\begin{equation}
= \bm \Omega \cdot \nabla \int_0^s \varphi (\bm x,s') ds' +  \frac{(1 - \varepsilon^2 \kappa) b_1}{1-(1 - \varepsilon^2 \kappa) b_1} \int_0^{\infty} p(s'') \left( \bm \Omega \cdot \nabla \int_0^{s''} \varphi (\bm x,s') ds' \right) ds''.
\end{equation}
For simplicity, we rename the coefficient
\begin{equation}
d_1 = \frac{(1 - \varepsilon^2 \kappa) b_1}{1-(1 - \varepsilon^2 \kappa) b_1}.
\end{equation}
Combining the results for $ i = 0 $ and $ i = 1 $, $ \Psi (\bm x, \bm \Omega, s) $ from Equation (59) becomes
\begin{equation}
\Psi (\bm x, \bm \Omega, s) = \frac{1}{4 \pi} \varphi (\bm x,s) \nonumber
\end{equation}
\begin{equation}
 - \frac{1}{4 \pi} \varepsilon \left[  \bm \Omega \cdot \nabla \int_0^s \varphi (\bm x,s') ds' + d_1 \bm \Omega \cdot \nabla \int_0^{\infty} p(s'') \left( \int_0^{s''} \varphi (\bm x,s') ds' \right) ds'' \right] + O \left( \varepsilon^2 \right).
\end{equation}
We now insert this result into the integral on the left side of Equation (27) in order to arrive at an equation in terms of only $ \varphi (\bm x,s) $.  Therefore, the term on the left side of Equation (27),
\begin{equation}
 \varepsilon \Im \bm \Omega \cdot \nabla \int_0^s \Psi (\bm x, \bm \Omega, s') ds',
\end{equation}
becomes
\begin{equation}
 \varepsilon \Im \bm \Omega \cdot \nabla \int_0^s \Psi (\bm x, \bm \Omega, s') ds' \nonumber
\end{equation}
\begin{equation}
= \frac{1}{4 \pi} \varepsilon \Im \bm \Omega \cdot \nabla \int_0^s \varphi (\bm x,s') ds' - \frac{1}{4 \pi} \varepsilon^2 \Im \bm \Omega \cdot \nabla \int_0^s \left[ \bm \Omega \cdot \nabla \int_0^{s''} \varphi (\bm x,s') ds' \right] ds''\nonumber
\end{equation}
\begin{equation}
-\frac{1}{4 \pi} d_1 \varepsilon^2 \Im \bm \Omega \cdot \nabla \int_0^s \left[ \bm \Omega \cdot \nabla \int_0^{\infty} p(s'') \left( \int_0^{s''} \varphi (\bm x,s') ds' \right) ds'' \right] ds''' + O \left( \varepsilon^3 \right).
\end{equation}
Next, the operator $ \Im $ is evaluated.  The first term in on the right side of Equation (90) equals zero since it is an odd function of $ \bm \Omega $.  The rest of Equation (90) becomes
\begin{equation}
 \varepsilon \Im \bm \Omega \cdot \nabla \int_0^s \Psi (\bm x, \bm \Omega, s') ds'  \nonumber
\end{equation}
\begin{equation}
= - \frac{1}{4 \pi} \frac{1}{3} \varepsilon^2 \left[ \nabla^2 \left( \int_0^s \varphi (\bm x,s) ds' \right)^2 \right] \nonumber
\end{equation}
\begin{equation}
- \frac{1}{4 \pi} \frac{1}{3} \varepsilon^2 \left[ d_1 \nabla^2 \int_0^s \left( \int_0^{\infty} p(s'') \left( \int_0^{s''} \varphi (\bm x,s') ds' \right) ds'' \right) ds''' \right] + O \left( \varepsilon^3 \right).
\end{equation}
We now insert this result into Equation (26).  This yields
\begin{equation}
\frac{\varphi (\bm x,s)}{4 \pi} - \frac{1}{4 \pi} \frac{1}{3} \varepsilon^2 \left[ \nabla^2 \left( \int_0^s \varphi (\bm x,s) ds' \right)^2 \right] \nonumber
\end{equation}
\begin{equation}
- \frac{1}{4 \pi} \frac{1}{3} \varepsilon^2 \left[ d_1 \nabla^2 \int_0^s \left( \int_0^{\infty} p(s'') \left( \int_0^{s''} \varphi (\bm x,s') ds' \right) ds'' \right) ds''' \right] + O \left( \varepsilon^3 \right) \nonumber
\end{equation}
\begin{equation}
 = \frac{1}{4 \pi} (1 - \varepsilon^2 \kappa)\int_0^\infty p(s') \varphi (\bm x,s')ds'  + \frac{1}{4\pi} \varepsilon^2 \langle s \rangle_\varepsilon q(\bm x).
\end{equation}
Simplifying to eliminate the $ 4 \pi $ in the denominator, this becomes
\begin{equation}
\varphi (\bm x,s) -  \frac{1}{3} \varepsilon^2 \left[ \nabla^2 \left( \int_0^s \varphi (\bm x,s) ds' \right)^2 \right] \nonumber
\end{equation}
\begin{equation}
- \frac{1}{3} \varepsilon^2 \left[ d_1 \nabla^2 \int_0^s \left( \int_0^{\infty} p(s'') \left( \int_0^{s''} \varphi (\bm x,s') ds' \right) ds'' \right) ds''' \right] + O \left( \varepsilon^3 \right) \nonumber
\end{equation}
\begin{equation}
 = (1 - \varepsilon^2 \kappa)\int_0^\infty p(s') \varphi (\bm x,s')ds'  + \varepsilon^2 \langle s \rangle_\varepsilon q(\bm x).
\end{equation}
Finally, We now rewrite this in operator form as 
\begin{equation}
\left(I-\frac{1}{3} \varepsilon^2 \left[ \nabla^2 \left( \int_0^s (\cdot) ds' \right)^2 + d_1 \nabla^2 \int_0^s \left( \int_0^{\infty} p(s'') \left( \int_0^{s''} (\cdot) ds' \right) ds'' \right) ds''' \right] + O \left( \varepsilon^3 \right) \right) \varphi (\bm x,s)\nonumber
\end{equation}
\begin{equation}
  = (1 - \varepsilon^2 \kappa)\int_0^\infty p(s') \varphi (\bm x,s')ds'  + \varepsilon^2 \langle s \rangle_\varepsilon q(\bm x).
\end{equation}

\subsection{Derivation of $ SP_1 $}

\hspace{0.25in}
We proceed to determine an expression for the scalar flux $ \Phi(\bm x) $, which will lead to the nonclassical $SP_1$ equation with anisotropic scattering \cite{larsen_5}.

Since the right-hand side of Equation (94) is only a function of $ \bm x $, the left-hand side must be a separable function of $ s $ and $ \bm x $.  Then we can write
\begin{equation}
\varphi(\bm x, s) = g(s) \phi(\bm x),
\end{equation}
where $ g(s) $ is a power series of $ s $ given by
\begin{equation}
g(s) = g_0(s) + \varepsilon^2 g_2(s) + O(\varepsilon^3).
\end{equation}
We express $ \phi(\bm x) $ as the power series
\begin{equation}
\phi(\bm x) = \phi_0(\bm x) + \varepsilon^2 \phi_2(\bm x) + O(\varepsilon^3).
\end{equation}
Then, truncating at $ O(\varepsilon^3) $, we express the left side of Equation (94) as
\begin{equation}
\left(I-\frac{1}{3} \varepsilon^2 \left[ \nabla^2 \left( \int_0^s (\cdot) ds' \right)^2 + d_1 \nabla^2 \int_0^s \left( \int_0^{\infty} p(s'') \left( \int_0^{s''} (\cdot) ds' \right) ds'' \right) ds''' \right] \right) \nonumber
\end{equation}
\begin{equation}
\cdot \left( Ig_0(s) + \varepsilon^2g_2(s) \right) \left( \phi_0(\bm x) + \varepsilon^2 \phi_2(\bm x) \right).
\end{equation}
Collecting terms by powers of $ \varepsilon $ yields
\begin{equation}
\varepsilon^0:  Ig_0(s)\phi_0(\bm x) \rightarrow g_0(s) = 1.
\end{equation}
\begin{equation}
\varepsilon^2:  \phi_2(\bm x) -  \frac{1}{3} \left[ \nabla^2 \left( \int_0^s (\cdot) ds' \right)^2 + d_1 \nabla^2 \int_0^s \left( \int_0^{\infty} p(s'') \left( \int_0^{s''} (\cdot) ds' \right) ds'' \right) ds''' \right] \phi_0(\bm x) \nonumber
\end{equation}
\begin{equation}
 + g_2(s) \phi_0(\bm x).
\end{equation}
Since the right side of Equation (94) is only a function of $ \bm x $, the coefficient in front of the $ \phi_0 (\bm x) $ term in Equation (100) must be zero, so
\begin{equation}
g_2(s) =  \frac{1}{3} \left[ \nabla^2 \left( \int_0^s (\cdot) ds' \right)^2 + d_1 \nabla^2 \int_0^s \left( \int_0^{\infty}  p(s'') \left( \int_0^{s''} (\cdot) ds' \right) ds'' \right) ds''' \right] \nonumber
\end{equation}
\begin{equation}
= \nabla^2 \frac{s^2}{2} + d_1 \nabla^2 \langle s \rangle_\varepsilon s.
\end{equation}
Then $ \varphi(\bm x, s) $ becomes
\begin{equation}
\varphi(\bm x, s) = \left( I + \varepsilon^2 \frac{1}{3} \left[ \nabla^2 \frac{s^2}{2} + d_1 \nabla^2 \langle s \rangle_\varepsilon s \right] \right) \phi(\bm x) + O(\varepsilon^3) .
\end{equation}
To get an expression for the scalar flux, which is 
\begin{equation}
\Phi(\bm x) = \int_{4\pi} \int_0^\infty \Psi(\bm x, \bm \Omega, s) \frac{e^{-\int_0^s \Sigma_t(s') ds'}}{\langle s \rangle_\varepsilon} ds d\Omega = \int_0^\infty \varphi(\bm x, s) \frac{e^{-\int_0^s \Sigma_t(s') ds'}}{\langle s \rangle_\varepsilon} ds ,
\end{equation}
we multiply Equation (102) by $ \frac{e^{-\int_0^s \Sigma_t(s') ds'}}{\langle s \rangle_\varepsilon} $ and then operate on it  by $ \int_0^{\infty}(\cdot) ds $.  This results in 
\begin{equation}
\Phi(\bm x) = \left( I + \varepsilon^2 \frac{1}{3} \left[ \nabla^2 \frac{\langle s^3 \rangle_\varepsilon}{3!\langle s \rangle_\varepsilon} + d_1 \nabla^2 \frac{\langle s^2 \rangle_\varepsilon}{2} \right] \right) \phi(\bm x) + O(\varepsilon^3) .
\end{equation}
Multiplying Equation (102) by the exponential term insures that when the equation is integrated from zero to infinity, the scalar flux will be finite.

Returning to Equation (94), we can express the integral on the right-hand side as a power series of the scalar flux as
\begin{equation}
\int_0^{\infty} p(s) \varphi(\bm x, s) ds = \left[ U_0 + \varepsilon^2 U_1 \nabla^2 + O(\varepsilon^3) \right] \Phi(\bm x).
\end{equation}
Incorporating Equation (102) into the left-hand side of Equation (105) results in
\begin{equation}
\int_0^{\infty} p(s) \varphi(\bm x, s) ds = \int_0^{\infty} p(s) \left[\left( I + \varepsilon^2 \frac{1}{3} \left[ \nabla^2 \frac{s^2}{2} + d_1 \nabla^2 \langle s \rangle_\varepsilon s \right] \right) \phi(\bm x) + O(\varepsilon^3) \right] ds.
\end{equation}
Integrating, this becomes
\begin{equation}
\int_0^{\infty} p(s) \varphi(\bm x, s) ds = \left[ 1 + \varepsilon^2 \frac{1}{3} \left( \nabla^2 \frac{\langle s^2 \rangle_\varepsilon}{2} + d_1 \nabla^2 \langle s \rangle_\varepsilon^2 \right) \right] \phi(\bm x) + O(\varepsilon^3).
\end{equation}
Inserting Equations (104) and (107) into Equation (105), excluding terms of $ O(\varepsilon^3) $ from both equations, and collecting powers of $ \varepsilon $ through equating both sides, we get
\begin{equation}
\varepsilon^0:  1 = U_0.
\end{equation}
\begin{equation}
\varepsilon^2:  \frac{1}{3} \left( \nabla^2 \frac{\langle s^2 \rangle_\varepsilon}{2} + d_1  \nabla^2 \langle s \rangle_\varepsilon^2 \right) = U_1 \nabla^2 + \frac{1}{3} \left( \nabla^2 \frac{\langle s^3 \rangle_\varepsilon}{3!\langle s \rangle_\varepsilon} + d_1  \nabla^2 \frac{\langle s^2 \rangle_\varepsilon}{2} \right) U_0.
\end{equation}
From this analysis, 
\begin{equation}
U_1 = \frac{1}{3} \left[ \frac{\langle s^2 \rangle_\varepsilon}{2} - \frac{\langle s^3 \rangle_\varepsilon}{3!\langle s \rangle_\varepsilon} + d_1 \left( \langle s \rangle_{\varepsilon}^2 - \frac{\langle s^2 \rangle_{\varepsilon}}{2} \right)  \right].
\end{equation}
We incorporate Equation (105) into Equation (94) and write the left-hand side of Equation (94) as a power series of the scalar flux.  This becomes
\begin{equation}
 \left[ V_0 + \varepsilon^2 V_1 \nabla^2 + O(\varepsilon^3) \right] \Phi(\bm x) = (1 - \varepsilon^2 \kappa) \left[ U_0 + \varepsilon^2 U_1 \nabla^2 + O(\varepsilon^3) \right] \Phi(\bm x) + \varepsilon^2 \langle s \rangle_\varepsilon q(\bm x).
\end{equation}
Using Equation (110), we can expand the series in both sides.  Once again, we exclude terms of $ O(\varepsilon^3) $, and we collect powers of $ \varepsilon $ on the left side to determine the $ V_n $ terms.  This results in
\begin{equation}
\varepsilon^0: I V_0 \phi_0(\bm x) \rightarrow V_0 = 1.
\end{equation}
\begin{equation}
\varepsilon^2:  \nabla^2 \phi_2(\bm x) + V_1 \nabla^2 \phi_0(\bm x) + \frac{1}{3} \left( \nabla^2 \frac{\langle s^3 \rangle_\varepsilon}{3!\langle s \rangle_\varepsilon} + d_1 \nabla^2 \frac{\langle s^2 \rangle_\varepsilon}{2} \right) \phi_0(\bm x).
\end{equation}
This implies that
\begin{equation}
V_1 = -\frac{1}{3} \left( \frac{\langle s^3 \rangle_\varepsilon}{3!\langle s \rangle_\varepsilon} + d_1 \frac{\langle s^2 \rangle_\varepsilon}{2} \right).
\end{equation}
Incorporating the expressions for $U$ and $V$ derived and simplifying, the truncated version of Equation (111) can be expressed as 
\begin{equation}
\left( W_1 \nabla^2 + \kappa U_0 \right) \Phi(\bm x) = \langle s \rangle_\varepsilon q(\bm x),
\end{equation}
where
\begin{equation}
W_1 = V_1 - U_1.
\end{equation}
Inserting the derived terms for $ W_1 $ and $ U_0 $, this becomes
\begin{equation}
\left( \left[ -\frac{1}{3} \left( \frac{\langle s^3 \rangle_\varepsilon}{3!\langle s \rangle_\varepsilon} + d_1  \frac{\langle s^2 \rangle_\varepsilon}{2} \right) \right] - \frac{1}{3} \left[ \frac{\langle s^2 \rangle_\varepsilon}{2} - \frac{\langle s^3 \rangle_\varepsilon}{3!\langle s \rangle_\varepsilon} + d_1  \left( \langle s \rangle_{\varepsilon}^2 - \frac{\langle s^2 \rangle_{\varepsilon}}{2} \right)  \right] \right) \nabla^2 \Phi(\bm x) \nonumber
\end{equation}
\begin{equation}
 + \kappa U_0 \Phi(\bm x)= \langle s \rangle_\varepsilon q(\bm x).
\end{equation}
Simplifying and rearranging terms results into a scaled diffusion equation for the scalar flux, and recalling that 
\begin{equation}
d_1 = \frac{(1 - \varepsilon^2 \kappa) b_1}{1-(1 - \varepsilon^2 \kappa) b_1}
\end{equation}
yields
\begin{equation}
-\frac{1}{3} \left[ \frac{\langle s^2 \rangle_\varepsilon}{2 \langle s \rangle_\varepsilon} +  \frac{(1 - \varepsilon^2 \kappa) b_1} {1-(1 - \varepsilon^2 \kappa) b_1} \langle s \rangle_\varepsilon \right] \nabla^2 \Phi(\bm x) + \frac{\kappa}{\langle s \rangle_\varepsilon} \Phi(\bm x) = q(\bm x).
\end{equation}
We now insert the scaling relationships
\begin{equation}
\langle s^2 \rangle_\varepsilon = \frac{\langle s^2 \rangle}{\varepsilon^2}, \hspace{0.1in} \langle s \rangle_\varepsilon = \frac{\langle s \rangle}{\varepsilon}, \hspace{0.1in} \kappa = \frac{1-c}{\varepsilon^2}, \hspace{0.1in} q(\bm x) = \frac{Q(\bm x)}{\varepsilon}
\end{equation}
into Equation (119) to arrive at an unscaled diffusion equation for the scalar flux, 
\begin{equation}
-\frac{1}{3} \left[ \frac{\langle s^2 \rangle}{2 \langle s \rangle} + \frac{c b_1}{1-c b_1} \langle s \rangle\right] \nabla^2 \Phi(\bm x) + \frac{1-c}{\langle s \rangle} \Phi(\bm x) = Q(\bm x).
\end{equation}
Recalling that  $ b_1 = \bar{\mu}_0 $, the nonclassical $SP_1$ equation with anisotropic scattering becomes \begin{equation}
-\frac{1}{3} \left[ \frac{\langle s^2 \rangle}{2 \langle s \rangle} + \frac{c \bar{\mu}_0}{1 - c \bar{\mu}_0} \langle s \rangle \right] \nabla^2 \Phi(\bm x) + \frac{1-c}{\langle s \rangle} \Phi(\bm x) = Q(\bm x).
\end{equation}
This agrees with the nonclassical $ SP_1 $ equation derived in \cite{larsen_5}.  If scattering is isotropic, then $\bar{\mu}_0 = 0$, and this result reduces to the equation nonclassical $SP_1$ equation with isotropic scattering given by Equation (12) \cite{vasques_4}, and if the scattering is anisotropic but the total macrocropic cross section is independent of the free-path length $s$, then Equation (122) reduces to the classical $SP_1$ equation with anisotropic scattering given by Equation (11) \cite{larsen_5}.
We note that this asymptotic analysis requires that the first two moments of $p(s)$ exist.  More generally, if $p(s)$ decays algebraically as $s \longrightarrow \infty$ such that \cite{vasques_4}
\begin{equation}
p(s) \geq \frac{constant}{s^{2N+1}} \hspace{0.1in} for s \gg 1,
\end{equation}
where $N$ is the same order as the nonclassical $SP_N$ equation, then \cite{vasques_4}
\begin{equation}
\langle s^{2N} \rangle = \int_0^\infty s^{2N} p(s) ds = \infty,
\end{equation}
and this asymptotic theory is invalid.

\subsection{Boundary Conditions}

\hspace{0.25in}
This asymptotic analysis does not produce boundary conditions, so we will show that the nonclassical $SP_1$ equation with anisotropic scattering given by Equation (122) can be manipulated into a classical form with modified parameters.  This will then allow us to use classical (Marshak) vacuum boundary conditions \cite{vasques_4}.  We define
\begin{equation}
\hat{\Sigma}_t = \left[ \frac{\langle s^2 \rangle}{2 \langle s \rangle} + \frac{c \bar{\mu}_0}{1 - c \bar{\mu}_0} \langle s \rangle \right]^{-1}
\end{equation}
and
\begin{equation}
\hat{\Sigma}_a = \frac{1-c}{\langle s \rangle}.
\end{equation}
Then, Equation (122) can be rewritten as
\begin{equation}
-\frac{1}{3 \hat{\Sigma}_t}  \nabla^2 \Phi(\bm x) + \hat{\Sigma}_a \Phi(\bm x) = Q(\bm x).
\end{equation}
Then the vacuum boundary conditions for Equation (122) are given by
\begin{equation}
\frac{1}{2} \Phi(\bm x) - \frac{1}{3 \hat{\Sigma}_t} \bm n \cdot \nabla \Phi(\bm x) = 0.
\end{equation}
We also note that if scattering is isotropic, this expression reduces to the vacuum boundary conditions used by Vasques and Slaybaugh \cite{vasques_4}, and if the total macroscopic cross section is independent of $s$, then this equation reduces to the classical Marshak vacuum boundary conditions.

\pagebreak

\section{Summary}

\hspace{0.25in}
This paper shows the development of a method which can be used to derive the nonclassical $SP_N$ equations with anisotropic scattering.  This procedure was used to derive the nonclassical $SP_1$ with anisotropic scattering, which was shown to be correct by reducing it to its nonclassical isotropic and classical anisotropic counterparts.  The nonclassical $SP_N$ equations with anisotropic scattering will be used to solve diffusive problems in which transport is nonclassical and scattering is anisotropic.  Since these equations are generalizations of their nonclassical isotropic and classical anisotropic counterparts, they will provide accurate solutions to diffusive problems in which the nature of the medium is less understood.  In future work, this method will be used to derive the higher order nonclassical $SP_N$ equations with anisotropic scattering, and then these equations will be validated numerically.

\pagebreak

\end{document}